\documentclass[useAMS,usenatbib]{mn2e}
\usepackage{graphicx} 
\usepackage{amsmath} 
\usepackage{xcolor}
\usepackage{caption}
\usepackage{placeins}
\usepackage{hyperref}
\usepackage{cleveref}
\usepackage{subcaption}
\usepackage{float}
\usepackage{lipsum}
\usepackage{multicol}
\usepackage{amssymb}
\def\refjnl#1{{\rm#1}}
\def\apj{\refjnl{ApJ}}                 % Astrophysical Journal
\def\apjl{\refjnl{ApJ}}                % Astrophysical Journal, Letters
\def\apjs{\refjnl{ApJS}}               % Astrophysical Journal, Supplement
\def\aap{\refjnl{A\&A}}                % Astronomy and Astrophysics
\def\mnras{\refjnl{MNRAS}}             % Monthly Notices of the RAS
\def\prd{\refjnl{Phys.~Rev.~D}}        % Physical Review D
\def\pasp{\refjnl{PASP}}               % Publications of the ASP
\def\nat{\refjnl{Nature}}              % Nature
\def\physrep{\refjnl{Phys.~Rep.}}   % Physics Reports
\def\be{\begin{equation}} 
\def\ee{\end{equation}}

\def\msun{{\Msun}}

\def\gsim{\lower.5ex\hbox{\gtsima}} 
\def\lsim{\lower.5ex\hbox{\ltsima}} \def\gtsima{$\; \buildrel > \over 
\sim \;$} \def\ltsima{$\; \buildrel < \over \sim \;$} \def\prosima{$\; 
\buildrel \propto \over \sim \;$} \def\gsim{\lower.5ex\hbox{\gtsima}} 
\def\lsim{\lower.5ex\hbox{\ltsima}} 
\def\simgt{\lower.5ex\hbox{\gtsima}} 
\def\simlt{\lower.5ex\hbox{\ltsima}} 
\def\simpr{\lower.5ex\hbox{\prosima}} \def\la{\lsim} \def\ga{\gsim} 
  
 \def\gtsima{$\; \buildrel > \over \sim \;$} 
\def\ltsima{$\; \buildrel < \over \sim \;$} 
\def\gsim{\lower.5ex\hbox{\gtsima}} 
\def\lsim{\lower.5ex\hbox{\ltsima}} 
\def\simgt{\lower.5ex\hbox{\gtsima}} 
\def\simlt{\lower.5ex\hbox{\ltsima}} 
\def\simpr{\lower.5ex\hbox{\prosima}} 
\def\la{\lsim} 
\def\ga{\gsim}

\def\msun{\,{\rm \Msun}}

\def\E3{{\cal E}_{\rm g}^{III}}

\def\Msun{\rm M_\odot}

\def\Zsun{\rm Z_\odot}

\def\Msun{\rm M_\odot}

\def\Zsun{\rm Z_\odot}
\def\M*{M_*}

\def\Z*{Z_*}
\def\L*{L_*}

\def\der{{\rm d}}

\def\der{{\rm d}}

\usepackage{hyperref} 
\title[PopIII SFRD]{A hint on the metal-free star formation rate density from 21cm-EDGES data} 
\author[Chatterjee et al.]{Atrideb Chatterjee$^{1}$\thanks{atrideb@ncra.tifr.res.in}, Pratika Dayal$^2$, Tirthankar Roy Choudhury$^1$,
\newauthor Raffaella Schneider$^{3,4,5}$\\ 
$^{{1}}$National Centre for Radio Astrophysics, Tata Institute of Fundamental Research, Pune 411007, India\\
$^{2}$Kapteyn Astronomical Institute, University of Groningen, P.O. Box 800, 9700 AV Groningen, The Netherlands\\
$^{3}$Dipartimento di Fisica, Sapienza, Universit$\grave{a}$ di Roma, Piazzale Aldo Moro 5, 00185, Roma, Italy  \\
$^{4}$INFN, Sezione di Roma I, P.le Aldo Moro 2, 00185 Roma, Italy  \\
$^{5}$INAF/Osservatorio Astronomico di Roma, Via di Frascati 33, 00078 Monte Porzio Catone, Italy  \\
 \\
}

\begin{document} 
 
\date{}

\maketitle 
 
\label{firstpage} 
\begin{abstract} 

We aim to provide here the first data-constrained estimate of the metal-free (Population III; Pop III) star formation rate density $\dot{\rho}_{*}^{III}$ required at high-redshifts ($z \gsim 16$) in order to reproduce both the amplitude and the redshift of the EDGES 21-cm global signal. Our model accounts for the Lyman Alpha (Ly$\alpha$), radio and X-ray backgrounds from both Pop III and metal-enriched Population II (Pop II) stars. For the latter, we use the star formation rate density estimates (and the Ly$\alpha$ background) from the {\it Delphi} semi-analytic model that has been shown to reproduce all key observables for galaxies at $z \gsim 5$; the radio and X-ray backgrounds are fixed using low-$z$ values. The constraints on the free parameters characterizing the properties of the Pop III stars are obtained using a Markov Chain Monte Carlo analysis. Our results yield a $\dot{\rho}_{*}^{III}$ that whilst increasing from $z \sim 21-16$ thereafter shows a sharp decline which is in excellent agreement with the results found by \citet{valiante2016} to simulate the growth of $z \sim 6 - 7$ quasars and their host galaxies, suggesting that the bulk of Pop III star formation occurs in the rarest and most massive metal-poor halos at $z \lesssim 20$. This allows Pop III stars to produce a rapidly growing  Ly$\alpha$ background between $z \sim 21-15$. Further,  
Pop III stars are required to provide a radio background that is about 
$3-4$ orders of magnitude
higher than that provided by Pop II stars although Pop II stars dominate the X-ray background.

\end{abstract}


%#################################################################
\section{Introduction}
%#################################################################

The past few years have seen enormous advances in peering back to the era of early star formation \citep[for a recent review see][]{dayal2018}, with instruments such as the Hubble Space telescope ({\it HST}) proving glimpses of star formation at redshifts as high as $z \sim 11$ \citep{oesch2016} with a resolution of a few parsecs \citep[e.g.][]{bouwens2017, vanzella2019}. However, the first metal-free (Population III; Pop III) stars, postulated to explain the metallicity-gap from Big Bang Nucleosynthesis (BBN) to the metal-rich (Population II; Pop II) stars seen in  high-redshift galaxies or in the low-metallicity tail of the metallicity distribution function of Galactic halo stars remain elusive. The fact that the first stars lasted only a few Myrs, if their initial mass function is predominantly top-heavy (see \citealt{bromm2013} for a thorough review) coupled with their predicted low star formation efficiency \citep{tornatore2007, maio2010, maio2011, johnson2013, xu2016b, debennassuti2017, jaacks2018, sarmento2019} has, so far, not resulted in the detection of any metal-free stars at high-$z$ (see however \citealt{vanzella2020} for a strongly lensed, candidate Pop III star complex at $z = 6.6$). Indeed, models require an instantaneous Pop III burst of $\sim 10^5 \msun$ for such first stars to be visible with the James Webb Space telescope \citep[JWST; e.g.][]{zackrisson2015}.    

However, recently, another avenue has opened up with regards to studying star formation at extremely high-redshifts ($z \gsim 16$). This is the result provided by the EDGES (Experiment to Detect the Global Epoch of Reionization Signature) collaboration \citep{bowman2018}. These authors have measured a sky-averaged 21cm absorption signal (from neutral hydrogen) at a central frequency $\nu = 78\pm 1$ MHz corresponding to $z \approx 17.2$. While this redshift is consistent with expectations of Lyman Alpha (Ly$\alpha$) photons coupling the spin temperature of neutral hydrogen to the gas kinetic temperature in the standard cold dark matter (CDM) paradigm, the differential brightness temperature $\sim -500\pm 200$mK is about twice as strong as that expected from any ``standard" model \citep{barkana2018}. Explaining the strength of this signal either requires gas colder than expected or a radiation background higher than expected \citep{bowman2018}. Indeed, given its possible origin from within the first 250 Myrs of the Universe, this signal has already been proposed to test new dark matter physics \citep{barkana2018, fraser2018, pospelov2018, slatyer2018, chatterjee2019}. On the other hand, the excess radio background solution \citep{feng2018, fialkov2019} could point to populations of early black holes/micro-quasars \citep{ewall_wice2018, mirabel2019}, Pop III stars \citep{jana2019, schauer2019} or extremely efficient star formation in low-mass halos \citep{mirocha2019}. 

In this work, we carry out a {\it proof-of-concept} calculation assuming that the EDGES signal is driven by a combination of Pop II and Pop III stars. Our aim is to provide the first data-constrained estimate of the star formation rate density (SFRD) at such high-$z$, using an MCMC approach, that results in a 21-cm signal {\it at the right redshift with an amplitude and shape compatible with the EDGES result}.  
 
 The cosmological parameters used in this work correspond to $\Omega_{\rm m } = 0.3089, \Omega_{\Lambda} = 0.6911, \Omega_{\rm b} = 0.049, h = 0.67, n_s = 0.96, \sigma_8 = 0.81$ \citep{planck2015}. 

%#################################################################
\section{Theoretical model}
\label{sec_model}
%#################################################################
The observable in global 21~cm experiments like EDGES is the globally averaged differential brightness temperature which, for the cosmological parameters used in this work, is given by \citep{chatterjee2019} 
\begin{equation}
\delta T_b(\nu) = 10.1~{\rm mK}~ \chi _{HI}(z)~\left(1 - \frac{T_{\gamma}(z)}{T_S(z)}\right)~(1 + z)^{1/2}
\end{equation}
where $\nu$ is the frequency of observation, $\chi_{HI}$ is the neutral Hydrogen fraction, $T_{\gamma}$ is the radiation temperature which includes the cosmic microwave background (CMB) plus a radio background, if any, and $T_S$ is the spin temperature of neutral hydrogen. In the above, we have assumed the intergalactic medium (IGM) to be predominantly neutral at redshifts corresponding to the EDGES observations.

We calculate the differential brightness temperature using the theoretical model detailed in \citet{chatterjee2019} which closely follows that of \citet{furlanetto2006c} and \citet{pritchard2012}. In brief, the spin temperature can be written as \citep{field1958}
\begin{equation}
    T_S^{-1}=\frac{T_{\gamma}^{-1}+x_cT_{k}^{-1}+x_{\alpha}T_\alpha^{-1}}{1+x_c+x_{\alpha}},
    \label{tspin}
\end{equation}
where $T_k$ is the gas kinetic temperature and $T_{\alpha}$ is the color temperature of the Ly$\alpha$ radiation field. Further, $x_c$ and $x_{\alpha}$ are the coupling coefficients corresponding to collisional excitations and spin-flip due to the Ly$\alpha$ radiation field, respectively. In this work, we assume $T_{\alpha}=T_k$. This is justified by the fact that the high optical depth for Ly$\alpha$ photons ensures that they undergo a large number of scatterings which are sufficient to bring the Ly$\alpha$ radiation field and the gas into local equilibrium near the central frequency of Ly$\alpha$ radiation \citep{pritchard2012}. 

We start by describing our calculation of the coupling coefficients before discussing the kinetic temperature in Eqn. \ref{tspin}. The collisional coupling coefficient, $x_c$, is determined by three different channels, namely, hydrogen-hydrogen (H-H), hydrogen-electron (H-e) and hydrogen-proton (H-p) collisions. 

Though the collisional coefficients do not play any important role at epochs relevant to this work we include them in our numerical code for completeness.

The Lyman Alpha (Ly$\alpha$) coupling coefficient, $x_{\alpha}$, is determined by the background Ly$\alpha$ flux, $J_{\alpha}$, as
\begin{equation}
x_{\alpha}=\frac{1.81 \times 10^{11}}{ (1+z)} S_{\alpha} \frac{J_{\alpha}} {\mbox{cm}^{-2} \mbox{s}^{-1} \mbox{Hz}^{-1} \mbox{sr}^{-1}},
\end{equation}
where $S_{\alpha}$ is a factor of order unity that accounts for the detailed atomic physics involved in the scattering process \citep{furlanetto2006b, hirata2006}. We calculate the total background Ly$\alpha$ flux at redshift $z$, accounting for the contribution from both Pop II and Pop III stars, as
\begin{equation}
J_{\alpha}(z) =\frac{c}{4 \pi}(1+z)^3 \int^{z_{max}}_{z}\dot{n}_{\nu'}(z') \left|\frac{d t'}{d z'}\right| dz'.
\end{equation}
Here $\nu' = \nu_{\alpha} (1 + z') / (1 + z)$, $\nu_{\alpha}$ is the Ly$\alpha$ frequency, $\dot{n}_{\nu'}(z')$ is the production rate of Ly$\alpha$ photons per unit frequency per unit comoving volume at redshift $z'$ which effectively scales with the star formation rate density, $c$ is the speed of light and $t'$ is the cosmic time corresponding to the redshift $z'$. The upper limit of the integral corresponds to $\nu' = \nu_{LL}$, i.e.,
\begin{equation}
1 + z_{max} = \frac{\nu_{LL}}{\nu_{\alpha}} (1 + z),
\end{equation}  
where $\nu_{LL}$ is the frequency corresponding to the Lyman-limit. Note that this upper limit implies that  $J_{\alpha}$ does not contain any contribution from ionizing photons $\nu \geq \nu_{LL}$ as they will be absorbed by neutral hydrogen atoms. For both Pop II and Pop III stars, we make the  assumptions that all photons with $\nu_{\alpha} < \nu < \nu_{LL}$ escape the host galaxy (since they avoid resonant scattering with neutral atoms) and that the spectrum of photons around the Ly$\alpha$ frequencies is constant \citep{furlanetto2006b}.

For Pop II stars, the rate $\dot{n}^{II}_{\nu'}(z)$ is directly obtained from the {\it Delphi} semi-analytic model. This value is obtained based on the entire assembly history of a galaxy using the {\it Starburst99} stellar population synthesis model \citep{leitherer1999} assuming a constant metallicity value of $5\%\Zsun$ and assuming each newly formed stellar population to have an age of 2 Myr. For simplicity, for Pop III stars, we assume $\dot{n}^{III}_{\nu'}(z')$ to be proportional to the corresponding star formation rate density $\dot{\rho}_{*}^{III}(z')$ such that
\begin{equation}
\frac{\dot{n}^{III}_{\nu'}(z')}{\rm{s^{-1}~Mpc^{-3}~Hz^{-1}}} = f_{\alpha}^{III} \frac{\dot{\rho}_{*}^{III}(z')}{\rm{\msun~Mpc^{-3}~yr^{-1}}},
\label{alpha3}
\end{equation}
where $f_{\alpha}^{III}$ is a proportionality factor that is treated as a free parameter and fixed using observations.

The gas kinetic temperature is determined by three processes: {\it (i)} the adiabatic cooling because of the expansion of the Universe, {\it (ii)} heating/cooling due to Compton scattering of residual electrons with background photons assuming an ionization fraction \citep[$\simeq 10^{-4}$;][]{bharadwaj2004}; and {\it (iii)} X-ray heating/cooling. For this process, the globally averaged energy injection rate per unit volume is taken to be proportional to the corresponding star formation rate density ($\dot{\rho}_*$) as 
\begin{align}
\frac{\epsilon_X}{\rm J~Mpc^{-3} ~s^{-1}} &= 3.4 \times 10^{33} \left(
f_{X,h}^{II} \frac{\dot{\rho}_{*}^{II}}{\rm{\msun~Mpc^{-3}~yr^{-1}}}
\right.
\notag \\
& + \left. f_{X,h}^{III} \frac{\dot{\rho}_{*}^{III}}{\rm{\msun~Mpc^{-3}~yr^{-1}}}
\right).
\end{align}
Here the proportionality factor $f_{X,h} = f_X \times f_h$, where $f_X$ is an unknown normalization factor allowing one to account for differences between the scaling obtained from local observations and that at high-$z$, and $f_h$ is the fraction of the X-rays that contribute to heating (the other part goes into ionization). The super-scripts denote the contributions of Pop II and Pop III stars. For Pop II stars, we assume that the scaling between the X-ray luminosity and the star formation rate remains unchanged as compared to low-$z$ observations and set $f^{II}_X = 1$. We also assume $f^{II}_h = 0.2$, appropriate for the early stages of reionization \citep{furlanetto2006c}. These assumptions fix the value of $f_{X,h}^{II} = 0.2$. For Pop III stars, we leave $f_{X,h}^{III}$ as a free parameter.

It has now been well established that in order to match the amplitude of the EDGES signal we would either require making the gas colder, using exotic physics such as new dark matter interactions \citep{barkana2018, fraser2018, pospelov2018, slatyer2018} or invoking a radiation temperature larger than the CMB temperature \citep{fraser2018,pospelov2018,ewall_wice2018,feng2018, jana2019}. In this work, we avoid incorporating any non-standard physics and focus on the latter scenario, as it has got some support by observations of an excess radio background in the local Universe, reported by the ARCADE-2 experiment \citep{fixsen2011}. We model the excess radio background by assuming that early galaxies produce radio frequency radiation whose strength is proportional to the star formation rate. The local radio-star formation rate ($L_{R}-\dot{M}_*$) relation at $150$ MHz is given by \citet{gurkan2018}
\begin{equation}
\frac{L_R}{\rm{J~s^{-1} Hz^{-1}}} = 10^{22} \left(\frac{\dot{M}_*}{\text{M}_{\odot}\text{yr}^{-1}}\right).
\end{equation}
We assume that a similar scaling holds at high redshifts too and hence the globally averaged radio luminosity per unit comoving volume (i.e., the radio emissivity) is given by
\begin{align}
\frac{\epsilon_R(z)}{{\rm J~s^{-1} Hz^{-1} Mpc^{-3}}} &= 10^{22} \left(
f_{R}^{II} \frac{\dot{\rho}_{*}^{II}}{\rm{\msun~Mpc^{-3}~yr^{-1}}}
\right.
\notag \\
& + \left. 
f_{R}^{III} \frac{\dot{\rho}_{*}^{III}}{\rm{\msun~Mpc^{-3}~yr^{-1}}}
\right)
\end{align}
where $f_R$ accounts for any differences between the local observations and high-$z$; we use $f_R^{II}=1$. The emissivity can be extrapolated from 150~MHz to higher frequencies by assuming a radio spectral index $\alpha_R$. The 21~cm radiation flux can then be written as \citep{2003ApJ...596....1C}
\begin{equation}
F_{R}(z)=  \left(\frac{1420}{150}\right)^{\alpha_R} \frac{c(1+z)^{3}}{4\pi} \int_z^{\infty} \epsilon_{R, \nu'}(z') \left|\frac{\der t'}{\der z'} \right| \der z',
\end{equation}
where $\epsilon_{R, \nu'}(z')$ is the comoving radio emissivity at $\nu' = 150~{\rm MHz} (1 + z') / (1 + z)$ and the factor $(1420/150)^{\alpha_R}$ on the right hand side accounts for the frequency-dependence of the radio emissivity. In this work, we assume $\alpha_R = -0.7$ \citep{gurkan2018}. Using the Rayleigh-Jeans law, the flux can be converted into a radio brightness temperature $T_{R}$, which results in a total background temperature given by $T_{\gamma}(z)=T_{R}(z)+T_{\rm CMB}(z)$. 
  
\begin{figure*}
\includegraphics[width=1.0\textwidth]{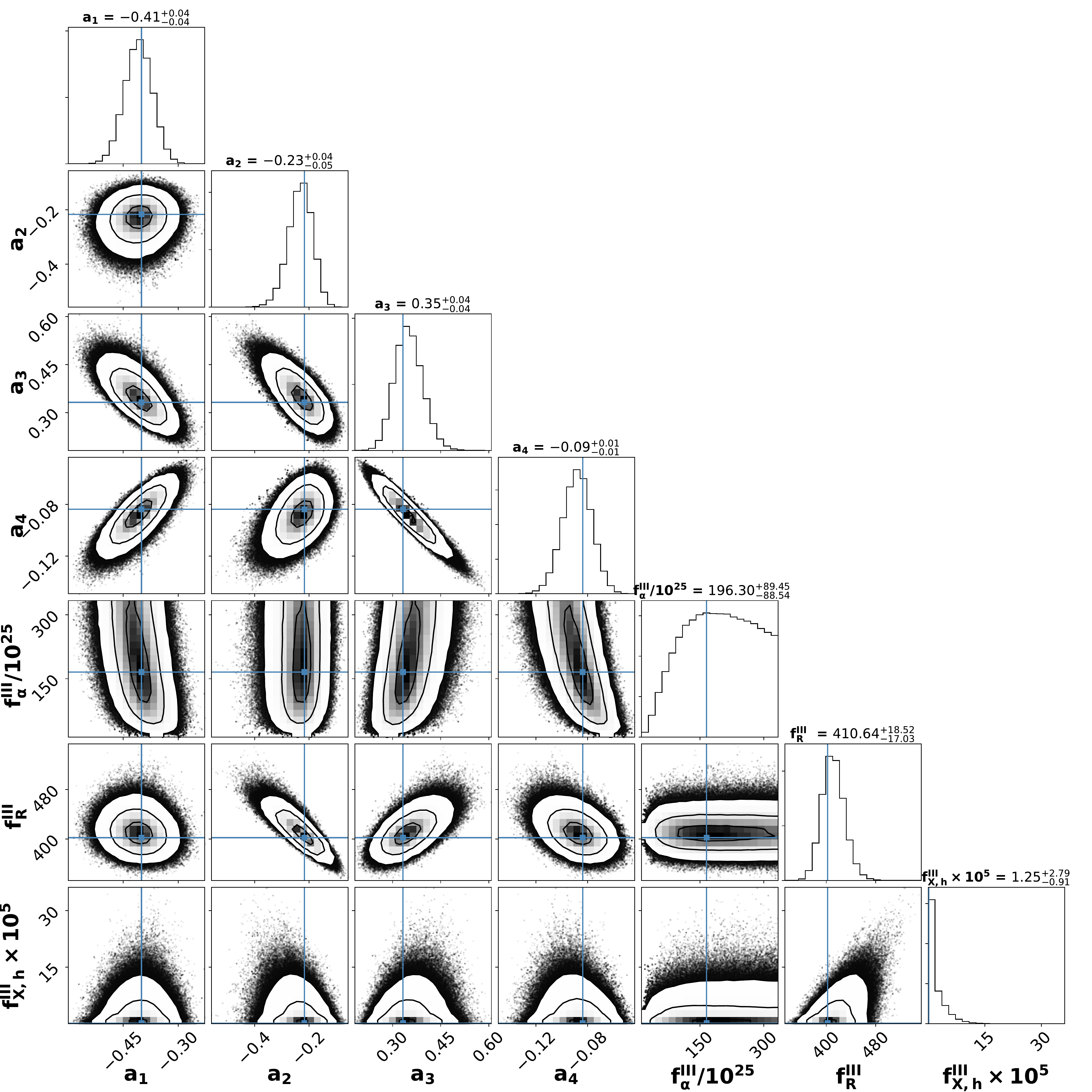}
\caption{The posterior distribution of the seven free parameters associated with the Pop III stars obtained by matching to the EDGES data (using the normalization $\dot{\rho}_{*}^{III}(z_{\rm mean})=10^{-4.5}~\Msun~{\rm yr^{-1}~Mpc^{-3}} $, see Sec. \ref{sec_model} for details). The two-dimensional plots show the joint probability distribution of any two parameters while the diagonal plots show the corresponding one-dimensional distributions. The values above the 1-D histograms show the median value along with the $\pm 1 \sigma$ range (i.e., the 16th and 84th percentiles). Contours in the 2-D plots correspond to the 1, 2, and $3\sigma$ ranges. The blue straight lines show the median values of the parameters.}
    \label{fig_corner}
\end{figure*}

\begin{figure*}
\includegraphics[width=1.0\textwidth]{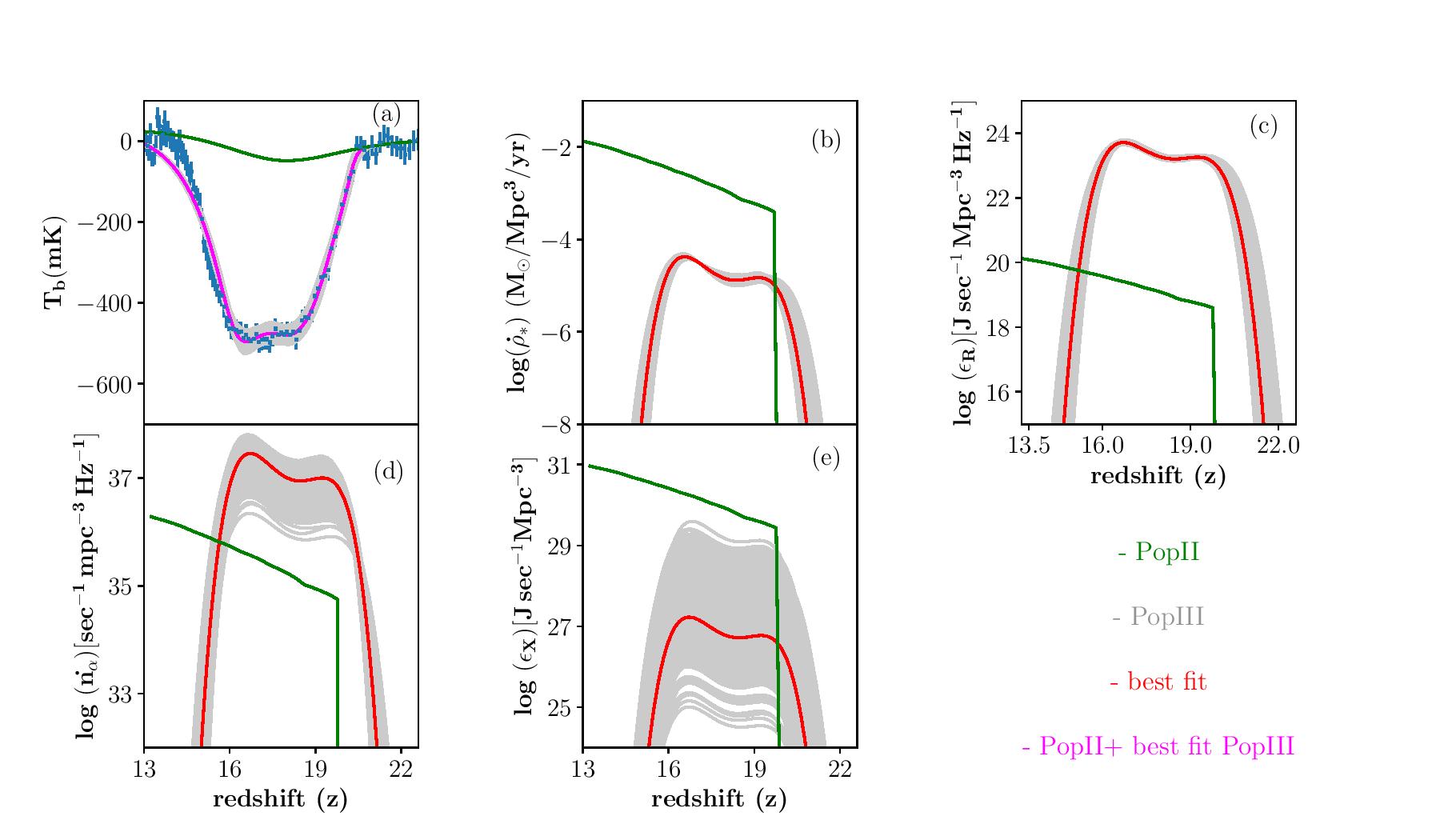}
\caption{As a function of redshift, the panels show the results for Pop II stars and Pop III stars for models that fit the EDGES brightness temperature for: {\it (a)}: the differential brightness temperature where the blue points show the EDGES data; {\it (b)}: the star formation rate densities; {\it (c)}: the radio emissivity; {\it (d)}: the Ly$\alpha$ photon production rate, and {\it (e)}: the X-ray emissivity. We artificially force the PopII SFRD to be zero at $z>19.5$ because the formation of PopII stars cannot precede that of PopIII stars. In all panels, the green and red lines show the results for Pop II (based on {\it Delphi}) and Pop III stars (the best-fit model obtained from MCMC) respectively. The grey shaded regions show 1000 random samples from the joint posterior distribution, i.e., they represent models that are consistent with the EDGES data.} 
    \label{fig_20mk}
\end{figure*}

As seen from this discussion, the only $z$-dependent quantities entering the equations explicitly are the star formation rate densities for Pop II and Pop III stars.  As already noted, for Pop II stars we use the star formation rate density and Ly$\alpha$ photon production rate results from the {\it Delphi} semi-analytic model that has been shown to reproduce all key observables for galaxies and AGN (including their UV luminosity functions, stellar mass/black hole mass densities, star formation rate densities, the stellar/black hole mass function) at $z \gsim 5$ \citep{dayal2014, dayal2019}. From this model, we find that the redshift evolution of the PopII star formation rate density is well described by $\log(\dot{\rho}_{*}^{II} / \Msun~Mpc^{-3}~\text{yr}^{-1})=-0.008(1+z)^2+0.047(1+z)-0.995$.

As for Pop III stars, we parametrize $\dot{\rho}_{*}^{III}(z)$ as function of redshift and attempt to constrain its shape using the EDGES observations. Note that the three essential physical quantities $\dot{n}_{\nu}$, $\epsilon_{X}$ and $F_{R}$ (or $T_R$) that determine the observable $\delta T_b(\nu)$ depend on the Pop III star formation rate density only through the combinations $f^{III}_{\alpha} \dot{\rho}_{*}^{III}$, $f^{III}_{X, h} \dot{\rho}_{*}^{III}$ and $f^{III}_{R} \dot{\rho}_{*}^{III}$, respectively. Hence the overall normalization of $\dot{\rho}_{*}^{III}(z)$ is \emph{completely degenerate} with the free parameters $f^{III}_{\alpha}$, $f^{III}_{X, h}$ and $f^{III}_{R}$. In view of this, we parameterize the Pop III star formation rate density using the log-polynomial form
\begin{align}
\log \left[\frac{\dot{\rho}_{*}^{III}(z)}{\dot{\rho}_{*}^{III}(z_{\rm mean})}\right]
& = \sum_{i=1}^4 a_i (z - z_{\rm mean})^i
\notag \\
&= a_1 (z-z_{\rm mean}) + a_2 (z-z_{\rm mean})^2 
\notag \\
& + a_3 (z-z_{\rm mean})^3 + a_4 (z-z_{\rm mean})^4
\label{eqn_sfrd3}
\end{align}
where $a_1$, $a_2$, $a_3$, $a_4$ are free parameters and $z_{\rm mean} = 17.15$ is the mean redshift of the EDGES observations. We find that parametrization of $\dot{\rho}_{*}^{III}(z)$ using a smaller number of parameters does not allow any good fit to the data, while using more terms in the polynomial leads to strong degeneracies between the coefficients. In order to simplify the problem, we use normalization such that $\dot{\rho}_{*}^{III}(z_{\rm mean}) = 10^{-4.5}~\Msun~{\rm yr^{-1}~Mpc^{-3}} $ which is motivated by the simulation results of \citet{valiante2016}. Note that this choice of the normalization has no impact on the constraints we obtain on the redshift-dependence of $\dot{n}_{\nu}$, $\epsilon_{X}$ and $F_{R}$ (or $T_R$) and the subsequent conclusions; all that would change when varying $\dot{\rho}_{*}^{III}(z_{\rm mean})$ would be the amplitude of the coefficients $f_{X,h}^{III}$, $f_{R}^{III}$, $f_{\alpha}^{III}$. We reiterate that, in our approach, the modelling of the Pop III star formation rate density is completely free of any physical assumptions and the form is solely constrained using data.

To summarize, our model has a total of seven free parameters: four for the Pop III star formation rate density ($a_1-a_4$) and three for the temperature contribution for Pop III stars ($f_{X,h}^{III}$, $f_{R}^{III}$, $f_{\alpha}^{III}$). Additionally, we have assumed $f_R^{II}=1$ and $f_{X,h}^{II}=0.2$. Finally, while the value of $\dot{\rho}_{*}^{II}$ is obtained from the {\it Delphi} semi-analytic model, we use a normalization of $\dot{\rho}_{*}^{III}(z_{\rm mean}) = 10^{-4.5}~\Msun~{\rm yr^{-1}~Mpc^{-3}}$.

By comparing the predictions of our model with EDGES data, we constrain the values of these parameters and consequently study the implications of the Pop III star formation at high-$z$. The constraints on the parameters, i.e., their posterior probability distributions are computed using a Bayesian statistical method based on MCMC. We make use of the publicly available {\tt emcee} package \citep{2013PASP..125..306F} for this purpose.

\begin{figure}
\includegraphics[width=0.5\textwidth]{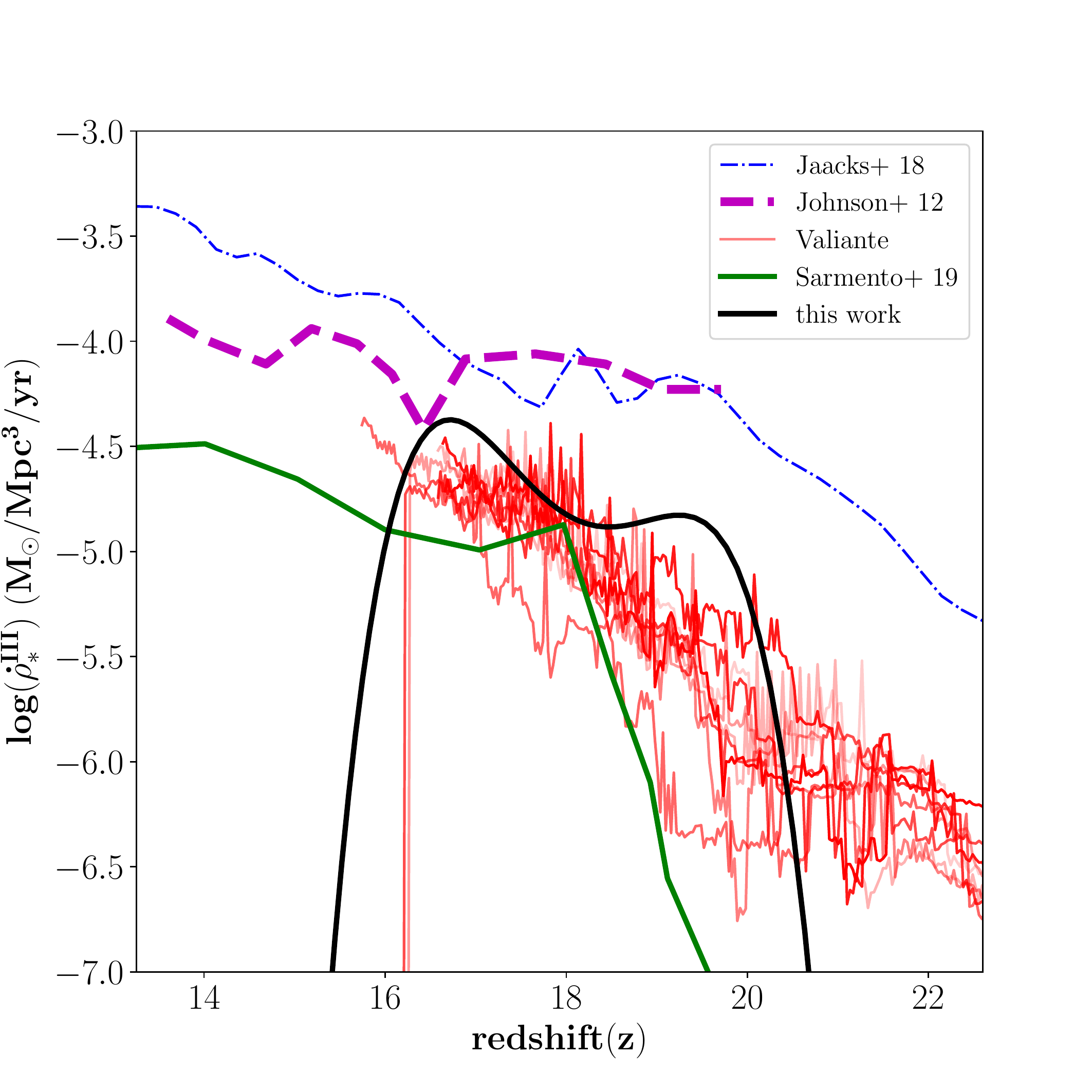}
\caption{The redshift evolution of the Pop III star formation rate density. The black line (best-fit model from this work) is compared to other theoretical estimates from: \citet[][dashed purple line]{johnson2013}, \citet[][dot-dashed blue line]{jaacks2018} and \citet[][solid green line]{sarmento2019}. Finally, the thin red lines show the results for ten different simulations by \citet{valiante2016}, that aim to reproduce the observed properties of $z \sim 6 - 7$ quasars and their host galaxies (see text).}
\label{fig_sfrd3}
\end{figure}

The MCMC analysis requires computing a likelihood ${\cal L} \propto {\rm e}^{-\chi^2 / 2}$, with the usual definition of the chi-square error such that
\begin{equation}
\chi^2 = \sum_i \left[\frac{\delta T_b^{\rm model}(\nu_i) - \delta T_b^{\rm data}(\nu_i)}{\sigma_i}\right]^2,
\end{equation}
where the sum is over all frequency bins corresponding to the EDGES data and $\sigma_i$ is the error. We take $\sigma_i = 25$mK, consistent with \citet{2018Natur.564E..32H}. We assume all the seven parameters to have broad flat priors: the priors for $a_1$, $a_2$, $a_3$, $a_4$ span the whole range from $-\infty$ to $\infty$ whereas $f^{III}_{\alpha}, f^{III}_{X, h}, f^{III}_{R}$ are allowed to vary between $0$ to $\infty$.   For exploring the parameter space with MCMC chains, we make use of 96 walkers\footnote{Essentially, walkers are the members of the ensemble which sample the posterior probability distribution function. They are similar to the popularly used Metropolis-Hastings chains, however, the proposal distribution for one walker is dependent on the positions of all the other walkers in the ensemble \citep{2013PASP..125..306F}} taking $1.5 \times 10^5$ steps. The initial $25\%$ steps are removed as ``burn-in'' and the remaining ones are used for estimating the posterior distributions. We have carried out the auto-correlation analysis \citep{Goodman2010} of the chains to ensure that the distributions have converged. 

%#################################################################
\section{Results}
%#################################################################
As already shown in \citet{chatterjee2019}, simultaneously matching to the shape and amplitude of the EDGES signal requires: a Ly$\alpha$ background already in place by $z \sim 20$ that can couple the spin temperature to the gas temperature, an excess radio background at $z \sim 20$ that yields a signal of the depth required and that falls off steeply at $z \lsim 17$ and an X-ray background by $z \sim 17$ that heats the gas so as to reduce the absorption amplitude. 

We start by quantifying the one- and two-dimensional projections of the posterior probability distributions of all the seven Pop III free-parameter values required to yield these trends in Fig. \ref{fig_corner}. Firstly, the increasing Pop III SFRD values at $z>z_{\rm mean}$ are essentially driven by the positive $a_3 (z-z_{\rm mean})^3$ term; the remaining three terms (with coefficients $a_1, a_2$ and $a_4$ in Eqn. \ref{eqn_sfrd3}) are responsible for the steep downturn seen at $z\lsim 16$. Moving to the fourth row of the same figure, we find $f_\alpha^{III} \sim 200 \times 10^{25}$. As seen in what follows, this results in a situation where Pop III stars dominate the Ly$\alpha$ background between $z \sim 21-15$. Further, as seen from these plots, per unit star formation rate density, Pop III stars are required to provide a radio background that is about 400 times higher that that provided by Pop II stars (where $f_R^{II}=1$), assuming the normalization $\dot{\rho}_{*}^{III}(z_{\rm mean}) = 10^{-4.5}~\Msun~{\rm yr^{-1}~Mpc^{-3}}$. This naturally results in a radio background that is predominantly driven by Pop III stars. Indeed, this is the key factor that determines the fall-off of the Pop III star formation rate density. Finally, we find that Pop II stars dominate the X-ray background at all $z$ with Pop III stars providing almost no contribution. This is required to maintain the spin temperature being coupled to the gas temperature between $z \sim 16-19$ and yield the EDGES absorption trough. Indeed, the earlier the X-ray background arises, the faster the spin temperature saturates to the CMB temperature.

Next, we focus on analyzing the contribution of Pop II and Pop III stars to the EDGES signal, as shown in Fig. \ref{fig_20mk}. In addition to the best-fit values for Pop III, we also show results for 1000 random points sampled from the chains; the region spanned by these sampled curves should be thought of as the range of histories that are allowed by the EDGES data. From panel (a), we see that our model results are in reasonable agreement with the EDGES data with the match being particularly good at $z \gtrsim 16$. We find that the match degrades at $z \lesssim 16$ where our models evolve somewhat slower than what is required by the data. This part of the evolution is driven by X-ray heating from Pop II stars, which in turn is fixed by the {\it Delphi} output; a better match would require steeper evolution in the Pop II star formation rate\footnote{It is
important to note that in {\it Delphi} all stellar populations are assumed to have a 5\% solar metallicity, so that all star forming halos at $z \leq 20$ are assumed to host Pop II stars. Hence, by using its prediction we are probably overestimating Pop II star formation rate at $z \sim 20 - 15$, since a fraction of star forming halos at these redshifts may still host Pop III stars.}.

Panel (b) in the same figure  shows the Pop III star formation rate density which is always subdominant compared to the Pop II stars. This is mainly because of the normalization we have chosen. However, as noted before, our key results concerns the {\it shape} of the Pop III star formation rate density. The data requires a rise in Pop III stars at $z \sim 20-21$ followed by a sharp fall at $z \sim 15-16$. The evolution of $\dot{\rho}_{*}^{III}(z)$ also shows a ``kink''-like feature at $z \sim 18$ which is due to a similar feature in the absorption profile as observed by EDGES. This is only to be expected - since our model of Pop III stars is entirely driven by data, any feature present in the data would show up in the model predictions as well. 
 
In panel (d), we see that the Ly$\alpha$ emissivity is completely dominated by Pop III stars at redshifts relevant for the EDGES observations. Note that, unlike $\dot{\rho}_{*}^{III}(z)$, this emissivity does \emph{not} depend on the normalization; this is because as seen from Eqn. \ref{alpha3}, the emissivity depends on the product of $\dot{\rho}_{*}^{III}(z)$ and $f_\alpha^{III}$ which absorb all the uncertainties. On the other hand, the X-ray emissivity from Pop III stars is always subdominant compared to the Pop II stars. This implies that the amount of heating required to match the EDGES data can be contributed solely by the PopII stars (as modeled in {\it Delphi}). Indeed, a larger contribution from Pop III stars would lead to a faster coupling between the spin and CMB temperatures, resulting in a higher redshift at which the signal would tend to 0. Finally, in all panels, we also show by the grey shaded regions the models that are consistent with the EDGES data. As expected, while the range of histories naturally show a dispersion, their {\it redshift-evolution} remains unaffected. The Pop III star formation rate density and the radio emissivity show the least dispersion (since they effectively govern the brightness temperature from Pop III stars) while, given their negligible contribution to the X-ray emissivity, the value of $f_X^{III}$ shows the largest dispersion. 
 
It is worth pointing out here that the depth of the EDGES signal is highly controversial \citep[see, e.g.,][]{2018Natur.564E..32H,Sims2020}. In case the amplitude of the signal turns out to be much smaller than what is assumed in this work (say $|\delta T_b| \lesssim 100$mK), then the Pop II stars with conventional choice of efficiency parameters are sufficient to explain the observations. Hence our conclusions on the Pop III stars hold \emph{only} in the case where the amplitude and redshift extent of the 21~cm signal are similar to those reported by the EDGES experiment.

Finally, we compare our predicted redshift evolution of the Pop  III star formation rate density with some theoretical models in Fig. \ref{fig_sfrd3}. Most models predict a Pop III star formation rate density that either rises between $z \sim 22 - 13$ \citep{jaacks2018, sarmento2019} or shows a relatively constant behaviour for $z \sim 20 - 13$ \citep{johnson2013}. The only exceptions are the predictions by \citet{valiante2016, valiante2018}, whose amplitude and redshift dependence are very similar to our predicted shape. Interestingly, the \citet{valiante2016, valiante2018} models aim to describe the formation histories of $z \sim 6 - 7$ quasars and their host galaxies. They therefore describe rapidly evolving and biased regions of the Universe. In these models, Pop III stars (and their black hole remnants) start to form at $z > 20$ in dark matter mini-halos with mass $[10^6 -  10^7] M_\odot$. However, their formation efficiency is small, as Lyman-Werner radiation limits gas cooling due to H$_2$ photo-dissociation. Because of this, the bulk of Pop III stars form at redshift $15 < z < 20$ in metal poor ($Z < Z_{\rm cr} =  10^{-3.8} Z_\odot$) Ly$\alpha$-cooling halos, with mass $[10^{7.5} - 10^{8.5}] M_\odot$ that are able to self-shield and cool the gas more  efficiently  (see the top panel  of Fig.1 in \citealt{valiante2018}). At $z \leq 15$ Pop III star formation is suppressed by metal enrichment, as $Z < Z_{\rm cr} =  10^{-3.8} Z_\odot$ star forming regions become increasingly rare. Hence, the shape of the EDGES signal suggests that the rapid coupling between the spin temperature and the gas kinetic temperature is driven by Ly$\alpha$ photons produced by Pop III stars that form in relatively massive and rare
halos, whose abundance grows fast,  due to the exponential growth of the massive-end of the halo mass function, before the quick rise of the gas temperature rapidly erases the signal at $z \lesssim 16$. Interestingly, these considerations are very consistent with the results of \citet{kaurov2018}, who also suggest that the shape of the EDGES signal can be best explained if the bulk of the UV photons are produced by rare and massive halos, and explore the implications for the amplitude of the 21-cm power spectrum.

%#################################################################
\section{Conclusions and discussion}
%#################################################################
In this work, we aim to provide an estimate of the Pop III star formation rate density required at high-redshifts ($z \gsim 16$) in order to reproduce both the amplitude and redshift of the EDGES 21-cm global signal. This essentially requires a rapidly rising Ly$\alpha$ background at $z \sim 20$ (to couple the spin temperature to the gas temperature), an excess radio background at $z \sim 20$ that falls off steeply at $z \lsim 17$ (to match to the depth and subsequent increase of the EDGES signal) and an X-ray background by $z \sim 17$ (to heat the gas so as to reduce the absorption amplitude). Our model has a total of seven free parameters: four for the Pop  III star formation rate density ($a_1-a_4$) and three for the temperature contribution for Pop III stars ($f_{X,h}^{III}$, $f_{R}^{III}$, $f_{\alpha}^{III}$); we use a normalization of $\dot{\rho}_{*}^{III}(z_{\rm mean}) = 10^{-4.5}~\Msun~{\rm yr^{-1}~Mpc^{-3}} $ motivated by simulation results of \citet{valiante2016}. These seven parameters are constrained using the {\tt emcee} package, a Bayesian statistical method based on MCMC. As for Pop II stars, the star formation rate density and Ly$\alpha$ photon production rate are obtained directly from the {\it Delphi} semi-analytic model that has been shown to reproduce all key observables for galaxies and AGN at $z \gsim 5$ \citep{dayal2014, dayal2019}. Finally, we use $f_R^{II}=1$ and $f_{X,h}^{II}=0.2$ in order to be consistent with low-redshift observations. {\it This results in a first estimate of the redshift evolution of the PopIII star formation rate density 
whose form is solely constrained using data}. 
Our key results are:
\begin{itemize}
\item Our model predicts a Pop III star formation rate density that whilst increasing from $z  \sim 21-16$ thereafter show a sharp decline. While this is in excellent agreement with the results of \citet{valiante2016, valiante2018}, there is a disparity with the results of most other models which predict a PopIII star formation rate density that either rises between $z \sim 22-13$ \citep{jaacks2018, sarmento2019} or shows a relatively constant behaviour for $z \sim 20-13$ \citep{johnson2013}.

\item Pop III stars dominate the Ly$\alpha$ background between $z \sim 21-15$.
\item Pop III stars are required to provide a radio background that is about  $3-4$ orders of magnitude higher than that provided by Pop II stars, resulting in a radio background that is predominantly driven by Pop III stars.
\item Pop II stars dominate the X-ray background at all $z$ (where $f_{X,h}^{II}=0.2$) with Pop III stars providing almost no contribution.

\end{itemize}

The above findings suggest that the rapidly rising Ly$\alpha$ background implied by the EDGES signal is dominated by the emission of Pop III stars formed in metal-poor Ly$\alpha$-cooling halos with mass $10^{7.5} - 10^{8.5} M_\odot$ \citep{valiante2016, valiante2018}. These systems are massive enough to provide a rapidly growing contribution at $z \sim 21 - 19$, due to the exponential growth of the massive end of the halo mass function (see also \citealt{kaurov2018}), but their number is large enough to ensure that the spin temperature and the gas kinetic temperature are coupled in a large fraction of the Universe
\footnote{Following \citet{kaurov2018}, the light travel distance between $z = 21$ and $z = 19$ can be estimated as $\sim 113$ h$^{-1}$ Mpc and the number density of the sources has to be $\geq 10^{-6.8}$ \, 
	(h$^{-1}$ Mpc)$^{-3}$, that is satisfied if the sources reside in halos with masses $\leq 10^{9.7} M_\odot$.}  In an earlier work, \cite{2020MNRAS.493.1217M} have studied the effects of Pop III remnants on the 21 cm signal and compared their results with the EDGES data. Analogous to this work, they have also found that the subset of their Pop III SFRD models that are able to match the timing and position of depth of the EDGES signal always prefer larger radio emission and small X-ray emission. In another recent (similar) work, using a more physically motivated models of Pop III stars, \citet{2019ApJ...877L...5S} find that Pop III stars are crucial in providing the necessary Ly$\alpha$-flux at high-redshifts in order to agree with the EDGES data, consistent with our conclusions regarding the Ly$\alpha$ flux. 
 
 We should also mention that our galaxy formation model (Delphi) and also the quantities, $f_{X}^{II}$ and $f_R^{II}$ for Pop II stars, are calibrated to lower redshift observations and could be different at higher redshift. However, in the absence of any rigorous model or observation at high redshifts, we have taken the most conservative approach and kept the values of these parameters constant. This also ensures that the number of free parameters do not increase beyond control
	
However, the question that remains is: how can Pop III stars dominate the Ly$\alpha$ and radio backgrounds and be completely subdominant for the X-ray background?
Pop III stars can give rise to a radio background as a result of synchrotron emission from high-energy cosmic rays (CR) electrons accelerated in SN explosions, but the signal is accompanied by substantial IGM heating due to CR protons \citep{jana2019}. Hence, even without including the effect of X-ray heating, the associated CR proton heating limits the contribution of Pop III SNe to the radio background required to explain the depth
of the absorption trough in the EDGES signal, unless the ratio of the CR electrons to protons energy is substantially larger than usually assumed. An additional contribution to the radio background at cosmic dawn could
come from accreting black hole remnants of Pop III stars. Again, to avoid the associated X-ray heating of the IGM, the X-rays must be absorbed by a substantial column of gas, with
$N_{\rm H} \sim 5 \times 10^{23}$ cm$^{-2}$, for $\sim 100$\, Myr \citep{ewall_wice2018}. This requirement allows to have escape fractions of ionizing photons $f_{\rm esc} \sim 0$ to avoid exceeding the Planck constraint
on the Thompson scattering optical depth \citep{Planck2016} and boosting-up the emission of Ly$\alpha$ photons. 

Whatever the true physical explanation is, it is clear that the shape of the 21-cm signal measured by EDGES has offered a first glimpse on the potential of 21-cm observations to constrain the properties of the Universe at cosmic dawn.

% ***************************************************************************
\section*{Acknowledgments} 
% ****************************************************************************
We warmly thank Rosa Valiante for providing her estimates of the Pop III star formation rate densities. AC and TRC acknowledge support of the Department of Atomic Energy, Government of India, under project no. 12-R\&D-TFR-5.02-0700.
PD acknowledges support from the European Research Council's starting grant ERC StG-717001 (``DELPHI"), from the NWO grant 016.VIDI.189.162 (``ODIN") and the European Commission's and University of Groningen's CO-FUND Rosalind Franklin program. PD also thanks La Sapienza for their hospitality and thanks the Sexten centre of Astrophysics where a huge part of this work was carried out.
RS acknowledges support from the Amaldi Research Center funded by the MIUR program ``Dipartimento di
Eccellenza" (CUP:B81I18001170001). 

% **************************************************************************
 
\bibliographystyle{mnras}
\bibliography{pop3}

\newpage 
\label{lastpage} 
\end{document}